\documentclass[sigconf]{acmart}

\usepackage{graphicx}
\usepackage{stfloats}
\usepackage{placeins}
\usepackage{amsmath}
\usepackage{url}
\usepackage{multirow}
\usepackage{array}
\usepackage{subfigure}
\usepackage{verbatim}
\usepackage{booktabs} 

\setcopyright{rightsretained}

\acmDOI{10.475/123_4}

\acmISBN{123-4567-24-567/08/06}

\acmConference[ND]{Technical Report}{February 2019}{Notre Dame, Indiana, USA} 
\acmYear{2019}
\copyrightyear{2019}

\acmPrice{15.00}

\begin{document}
\title{EyeDoc: Documentation Navigation with Eye Tracking}

\author{Robert Wallace and Collin McMillan}
\affiliation{%
  \institution{University of Notre Dame}
  \streetaddress{Department of Computer Science and Engineering}
  \city{Notre Dame} 
  \state{IN} 
  \postcode{46556}
}
\email{cmc@nd.edu}


\begin{abstract}
We demonstrate EyeDoc, a tool for navigating software documentation with the use of the eyes.  When programming, developers often have many windows open such as an IDE, consoles and GUIs for third-party utilities, the application under development, and a web browser for navigating documentation.  Several studies have shown that the navigation among these different tasks imposes a small mental load which, over time, adds to a significant decrease in productivity.  One solution to this problem is to increase ``screen real estate'' with larger monitors and higher resolutions, so that more information can be presented to the programmer at one time.  But this solution has limits: the complexity and size of software is also increasing rapidly.  In this paper, we use eye-tracking technology to build a tool for navigating documentation.  All a programmer needs to do to use EyeDoc is look at the monitor containing the documentation.  The tool detects when the eyes glance at different components of the documentation, and allows the programmer to navigate to those components by gazes and blinking.  The programmer does not need to move his or her hands, or risk losing the context of other tasks on the desktop.  We built EyeDoc as a research prototype and proof-of-concept using low-cost consumer eye-tracking hardware and our own software built as a JavaScript addition to JavaDocs.  This paper describes the tool's design, use, and strategy for evaluation and future development.
\end{abstract}

%
%



\maketitle

\vspace{-0.2cm}

\section{Introduction}
\label{sec:introduction}

The term ``context switch'' refers to the cost imposed when switching from one task to another.  Readers in Computer Science are likely familiar with the concept in operating systems process management, but the principle is also widely studied and documented in Psychology: many studies show how a context switch during a complex task reduces a human's performance on that task~\cite{rosas2006context, gamez2017roles, abad2009partial}, even when the human is not aware of the effects of the switch~\cite{diede2017cognitive}.  In Program Comprehension, a context switch often occurs during navigation of source code and documentation.  While a programmer has an IDE open and is reading code, he or she may have a question about the code which can be answered by reading documentation.  A context switch occurs as the programmer moves his or her attention away from the code, hands away from the keyboard and mouse, and documentation replaces code and other utilities on the screen.  The more time the switch takes, the more information the programmer loses, and the greater cognitive distance the programmer will have to travel when returning to the code.  As literature in both Software Engineering and Psychology has pointed out~\cite{jin2009self, perry1995understanding, robillard2004effective}, each switch imposes a small cost which over time adds to a measurable productivity penalty.

Different strategies to reduce the cost of the context switch have been implemented.  These strategies range from improved IDE designs, popup information boxes such as tooltips, autocompletion, and the use of larger and/or multiple monitors.  These strategies have been quite effective, but they still depend on the traditional keyboard-mouse interface.  However, recent advances in eye-tracking technology have reduced both the price and size of eye-tracking hardware, and offer the possibility of augmenting keyboard-mouse navigation with navigation based on the movement of the eyes~\cite{duchowski2007eye}.

In this paper, we demonstrate EyeDoc, a tool that allows programmers to navigate documentation by moving only their eyes.  The intent of the tool is to reduce the cognitive cost of the context switch from code to documentation, by reducing the time and number of steps required to complete the switch.  Instead of moving a hand to the mouse or completing one or more keyboard shortcuts, the programmer only needs to glance at the documentation to navigate it.  Productivity savings are possible because the programmer can return to the code context from the documentation context without even losing his or her cursor position.

The idea of navigation based on eye movements is not new: it has long been a component of assistive technologies for persons with motor impairments~\cite{istance1996providing}, and is supported by several eye-tracking hardware products as a feature called ``active eye tracking''~\cite{ooms2015accuracy}.  However, EyeDoc is a novel application of the technology in Program Comprehension, and is part of a broader trend towards programming interfaces that respond automatically to programmer behavior~\cite{robillard2017demand, kevic2017eye} and reduce interruptions~\cite{zuger2017reducing}.  We describe the design and implementation of the tool, how to use it, and our planned evaluation and key research questions.

\section{EyeDoc in a Nutshell}

\begin{figure}[!t]
	\begin{center}
		\includegraphics[width=8cm]{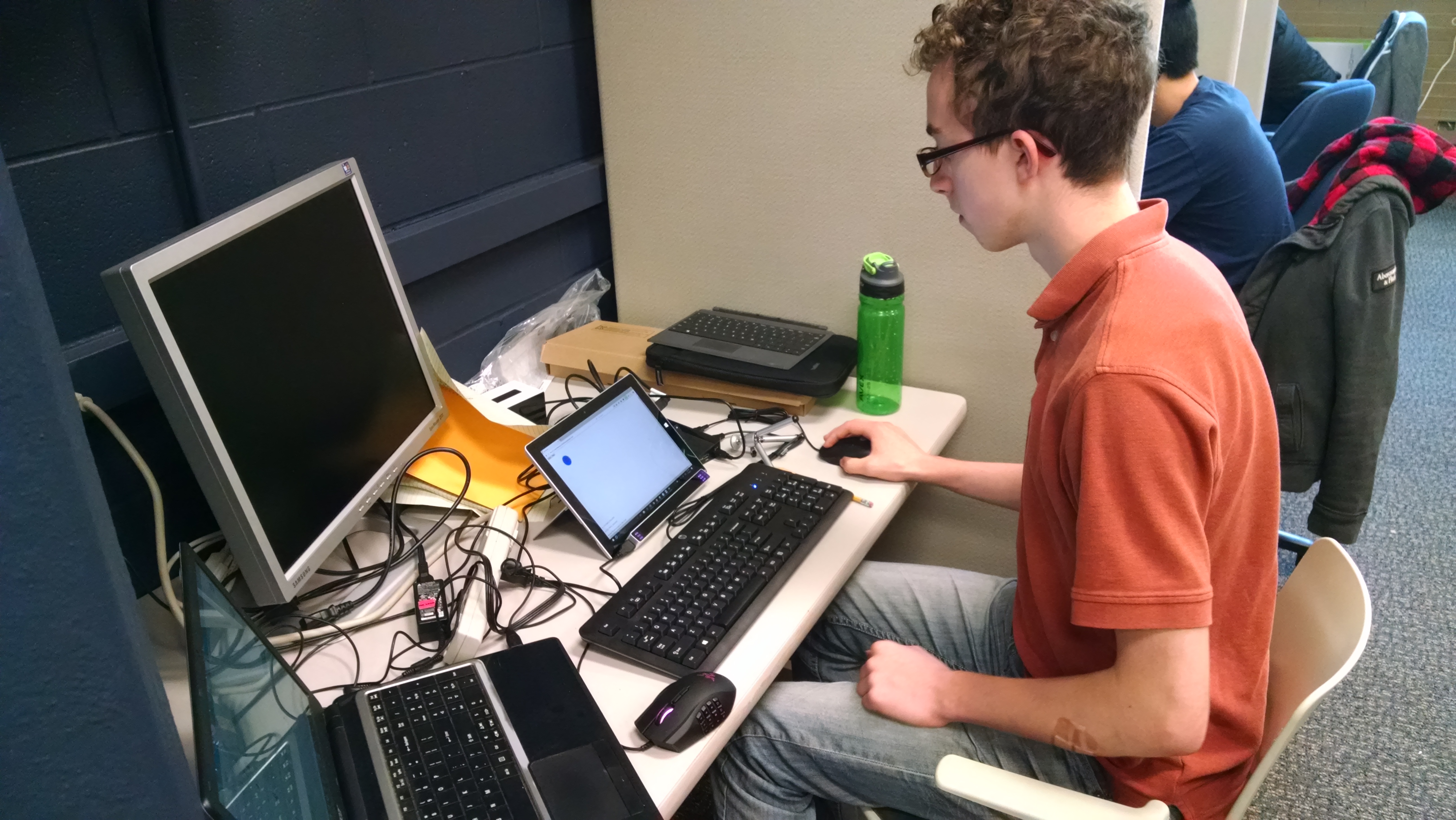}
		\vspace{-0.1cm}
		\caption{A deployment of EyeDoc.  The monitor to the left shows the programming environment.  The smaller monitor to the right shows EyeDoc.  The two monitors are adjacent, but to minimize setup costs and avoid conflicting with the development environment, EyeDoc is installed on a separate machine independent of the development machine.}
		\label{fig:eyedoc_setup}
	\end{center}
\end{figure}

In a nutshell and from a user's perspective, EyeDoc is a tool for navigating API documentation.  The prototype implementation we built is an interface that allows programmers to navigate that documentation by using eye movements and gazes, instead of the traditional keyboard and mouse.  We built EyeDoc as a proof-of-concept for eye-driven navigation during software development.  It has two components.  First is a monitor area dedicating to showing the EyeDoc visual interface, and second is the eyetracking hardware.  Figure~\ref{fig:eyedoc_setup} shows what we conceive of as a typical setup.  On the left is a large monitor with the programming IDE and code.  That large monitor is connected to a computer which is controlled by the keyboard and mouse.  On the right is a smaller monitor with the EyeDoc interface.  Below the smaller monitor is the eyetracking hardware.  EyeDoc operates independently from the IDE, and the user controls the interface entirely with the use of the eyes.  Therefore, EyeDoc may run on a separate computer as the IDE, to avoid dependencies and driver conflicts, should they arise.  In the image, EyeDoc is running on a dedicated Microsoft Surface 3 tablet computer, and the large monitor with the IDE is connected to a dedicated development laptop.  When the user wants to navigate the documentation, he or she only needs to move his or her eyes to EyeDoc.  The programmer can then read the documentation, and navigate by selecting links and navigation buttons by either looking at them and blinking, or gazing at them for a period of milliseconds (the navigation style and millisecond gaze delay is configurable).

\section{Envisioned Users}

We envision the users who would receive the most benefit from EyeDoc are programmers who frequently move between source code and documentation.  The more often that programmers must switch between code and documentation, the more opportunity exists for EyeDoc to reduce the cost of the switch.  While the relevant literature has long described documentation as one of the most common sources of knowledge for programmers~\cite{letovsky1987cognitive}, people who must use large and complex APIs, or a large number of smaller APIs, are more likely to need to make more context switches from code to documentation.  At the same time, EyeDoc has applications as an accessibility technology for persons with motor impairments, for whom the cost of moving the hands may add significant time to the context switch.  In that case, a benefit may be noticeable even when only a relatively few context switches are necessary.

\begin{figure}[!t]
	\begin{center}
		\includegraphics[width=8cm]{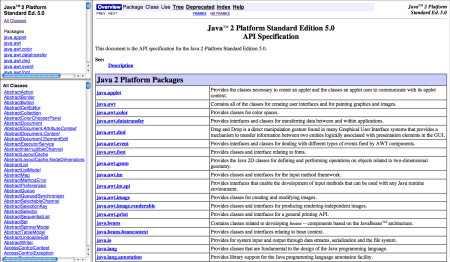}
		\vspace{-0.1cm}
		\caption{The interface displayed by the research prototype implementation of EyeDoc.  The tool automatically detects when the user looks at the interface, and navigates based on eye movements that are configurable.}
		\label{fig:eyedoc_interface}
		\vspace{-0.2cm}
	\end{center}
\end{figure}

\section{Tool Design and Implementation}
\label{sec:design}

\begin{figure}[!b]
	\begin{center}
		\vspace{-0.15cm}
		\includegraphics[width=6cm]{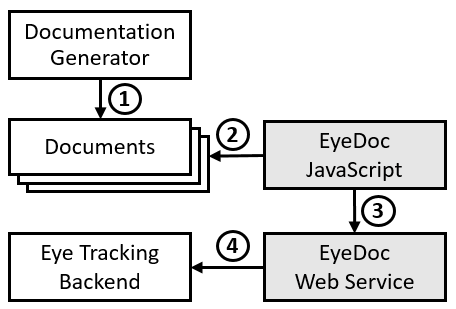}
		\vspace{-0.15cm}
		\caption{The architecture of EyeDoc at a high level.  Essentially, EyeDoc consists of a JavaScript front-end that is inserted into HTML documents.  The front-end then communicates with a web service, which provides access to a lower-level backend for the eye tracking hardware.  Section~\ref{sec:design} gives more details and implementation specifics.  Shaded areas indicate components we built, while unshaded areas indicate dependencies and third-party components.}
		\label{fig:eyedoc_arch}
	\end{center}
\end{figure}

We designed EyeDoc using a typical web architecture, with a JavaSc-ript addition to HTML documentation and a web service and hardware interface backend.  An alternative was to build EyeDoc as a browser plugin or standalone program, given that our tool runs on a local hardware host, but we decided in favor of a web architecture in order to maximize compatibility with HTML documents and browsers.  Figure~\ref{fig:eyedoc_arch} shows the architecture.  First, a third-party documentation generator parses the source code and creates documentation formatted as HTML (area 1).  Then, we inject our JavaScript frontend component into the \texttt{head} section of the HTML documents (area 2).  We provide a small script for this purpose as noted in the previous section.  When the programmer opens the documentation, the EyeDoc JavaScript will activate, and communicate with a web service backend for updates on eye movements (area 3).  The JavaScript continuously polls the web service even when the programmer looks away from the eye tracking screen area, so that it will be ready for navigation as soon as the programmer looks back at the screen area.  The web service is built on top of a third-party backend that communicates with and manages the eye tracking hardware (area 4).  The web service provides an abstraction layer between the JavaScript and the hardware interface, so that the frontend should not need to be altered for hardware upgrades.

Our implementation is based on the popular JavaDoc~\cite{kramer1999api} documentation generator.  In principle, EyeDoc could function on any HTML documentation (e.g. from Doxygen or other tools) with minor changes to how it detects API components, descriptions, and navigation areas (such as scroll buttons) in those files.  We chose JavaDoc due to its popularity and the uniformity of the files it generates.  We use Firefox 56.0.2 on Windows 10 as a test environment.  We built the web service using the CherryPy v3 Python framework.  The eye tracking hardware we used was the EyeTribe~\cite{ooms2015accuracy} with Java API version 0.9.77.  We hosted the software and hardware on a Microsoft Surface 3 tablet computer.  In theory, the web service and frontend could run on separate machines, though we execute them both on the tablet to avoid network latency.

\section{Evaluation Procedures}

We have planned an evaluation procedure to answer the following Research Questions (RQs):

\vspace{-0.1cm}
\begin{itemize}
	\item[RQ$_1$] Are programmers able to complete programming tasks more quickly when using EyeDoc than without?
	\item[RQ$_2$] Does EyeDoc reduce the time required for a context switch between reading source code and reading documentation?
	\item[RQ$_3$] Do programmers perceive a reduced workload or cognitive cost to reading documentation with EyeDoc than without?
\end{itemize}
\vspace{-0.1cm}

The rationale behind RQ$_1$ is to test whether programmers receive a benefit from using EyeDoc in terms of total time taken to complete a programming task.  This total time benefit is likely to be small, however, so we ask RQ$_2$ to measure the effect of EyeDoc on the context switches that programmers make, which is where we expect EyeDoc to provide the most value.  Finally, even if there are time benefits, it is possible that programmers will not perceive these benefits in the form of reduced workload, so we ask RQ$_3$ to test these perceptions.

Our planned methodology to answer these RQs is to recruit professional programmers to complete programming tasks with and without EyeDoc.  We aim to hire 30 professionals for two hour work sessions, paying the market rate in our area of US\$66/hr.  We will then use a cross-validation study design in which we rotate the tasks that we ask participants to perform and the order of the tool usage (i.e. some programmers will use EyeDoc first, while others will use it second).  We will set up the development environment to record both IDE actions, screen video capture, and eye movements.  The programming tasks will be such that most programmers should be able to complete 2-4 in a two hour period (ideally, at least one with EyeDoc and one without).  We will answer RQ$_1$ by measuring the total time per task, RQ$_2$ by measuring the time taken when switching between code and documentation, and RQ$_3$ with questionnaires during the study asking about perceived workload.  We feel this scope of the study is achievable based on the second author's experience conducting similar studies~\cite{rodeghero2015eye, armaly2017comparison, mcburney2016automatic, rodeghero2017detecting}.

\section{Related Work}

EyeDoc's related work can be broadly categorized as 1) active eye tracking interfaces, or 2) passive eye tracking experiments in software engineering.

Active eye tracking is the use of eye movements to facilitate navigation of computer interfaces.  The primary application area has traditionally been in assistive technologies for persons with motor impairments~\cite{istance1996providing}.  Examples have been well-documented by Majaranta~\cite{majaranta2011gaze} and Tai~\emph{et al.}~\cite{tai2008review}, and include efforts to help motor impaired children to draw, control of a movement apparatus such as a powered vehicle, control of speech synthesis systems, and control of GUIs for general computer use.  Active eye tracking for general computer use for all persons, regardless of disability status, has been proposed for many years~\cite{van1997post}, but has only recently gained traction as improvements in technology have made reliable eye tracking affordable.  Examples include smartphone navigation~\cite{heryadi2017mata}, entertainment games~\cite{tobiigames}, and GUI context detection to distinguish multiple simultaneous users of one computer~\cite{smith2017multi}.

Passive eye tracking experiments are experiments in which eye movements are tracked and recorded, but the movements do not affect the actions of the computer as perceived by the human user.  These experiments are effective at determining what information humans need, the order in which people search for information, and in detecting human factors that affect performance such as fatigue.  Due to space limitations, we direct readers to work by Sharif~\emph{et al.}~\cite{sharif2011use} and Rodeghero~\emph{et al.}~\cite{rodeghero2015eye} for work on passive experiments in SE.  Recently, hybrid passive and active eye tracking uses have been proposed to assist programmers in software engineering tasks, such as work by Shaffer~\emph{et al.}~\cite{shaffer2015itrace} and Kevic~\emph{et al.}~\cite{kevic2017eye}.

\vspace{-0.2cm}
\section{Limitations and Future Work}

As an early prototype, we view EyeDoc as a proof-of-concept and platform for future work.  The current prototype has a few technical limitations which we are addressing in ongoing work and deserve mention here: First, the current approach is deployable only for documents formatted as HTML.  While this encompasses a large number of popular API documentation generation systems (e.g. JavaDocs, Doxygen), we aim to make the approach functional on documentation embedded in an IDE or other locations.  Another limitation is that the JavaScript must be placed in every HTML page to be navigated.  In principle, a browser plugin would be more general and allow navigation of web sources, so the documentation does not need to be downloaded and modified locally prior to navigation.  Finally, EyeDoc requires significant computational resources to process the eye movements at high speed (a laptop is sufficient, but low-power devices such as tablets are generally not sufficient).  Much of this effort is overhead between the JavaScript and the web service, and is one target of our current technical work.

A trend in software documentation is towards on-demand documentation~\cite{robillard2017demand}, meaning documents that are responsive to programmers' needs at any particular moment.  At this time, EyeDoc is a tool for navigating static documentation.  However, our vision to use the eye tracking information to learn more about what the programmers' needs are, to help automatically detect those needs on the fly.  As both Sharif~\emph{et al.}~\cite{sharif2011use} and Rodeghero~\emph{et al.}~\cite{rodeghero2015eye} have pointed out, there are patterns to how programmers read that are likely to be different for different tasks.  EyeDoc could hypothetically detect these patterns and, in combination with other data sources and techniques, find and present information relevant to a programmers' immediate needs.  This research agenda connects to our current direction on automatic documentation generation~\cite{mcburney2016automatic} and automatic comprehension of programmer behavior~\cite{rodeghero2015eye}.

\vspace{-0.2cm}
\section{Conclusion}

We have demonstrated EyeDoc, a tool for navigating software documentation with the use of the eyes.  EyeDoc is intended to reduce the cost of a context switch between source code and documentation during program comprehension and other software engineering tasks.  We have built a prototype of EyeDoc that allows navigation of HTML-formatted API documentation such as JavaDocs.  Our implementation is built in a JavaScript frontend / web service backend architecture for reliable operation and use across browser versions.  We have released our implementation with a guide for setup and installation.  Finally, we presented our plans for evaluation and how EyeDoc will serve as a platform for future work.  Our hope for this demonstration is to disseminate knowledge we have gained, and to receive feedback on our current work to improve our future work.

\section{Acknowledgments}
This work is supported in part by the NSF CCF-1452959, CNS-1510329, and CCF-1717607 grants. Any opinions, findings, and conclusions expressed herein are the authors' and do not necessarily reflect those of the sponsors.

\bibliographystyle{ACM-Reference-Format}
\bibliography{biblio} 

\end{document}